\documentclass[10pt, a4paper, preprint]{article}
\usepackage[utf8]{inputenc}
\usepackage{multirow}
\usepackage[T1]{fontenc}
\usepackage{threeparttable} 
\usepackage{graphicx}
\usepackage{authblk}
\usepackage{natbib}

\title{Positron Emission Tomography with quantum-entangled Compton events: first imaging results at clinically relevant activities}

\author[1]{Mihael Makek}
\author[1,2,*]{Ana Marija Kožuljević}
\author[1]{Tomislav Bokulić}
\author[3]{Darko Grošev}
\author[4]{Zdenka Kuncic}
\author[5]{Siddharth Parashari}
\author[2]{Luka Pavelić}

\affil[1]{University of Zagreb Faculty of Science, Department of Physics, Zagreb, 10000, Croatia}
\affil[2]{Institute for Medical Research and Occupational Health, Division of Radiation Protection, Zagreb, 10000, Croatia}
\affil[3]{University Hospital Centre Zagreb, Department of Nuclear Medicine, Zagreb, 10000, Croatia}
\affil[4]{School of Physics, University of Sydney, Sydney, NSW 2006, Australia}
\affil[5]{Instituto de Física Corpuscular (IFIC), CSIC-UV, C/ Catedrático José Beltrán, 2, Paterna, Spain}

\affil[*]{akozuljevic@imi.hr}

\begin{document}

\maketitle
\begin{abstract}
In Positron Emission Tomography, a potential, yet unutilized enhancement may come from exploiting the quantum entanglement of the annihilation quanta, inscribed in the correlation of their polarizations. To investigate this, we built a PET demonstrator capable of measuring polarization correlations of annihilation quanta by their Compton scattering, based on single-layer scintillator polarimeters. We present a detailed study of the imaging of two $^{68}$Ge line sources, 45 MBq each, to extract the spatial resolution and assess image quality. The results show that a spatial resolution of 2.5$\pm$0.1 mm is obtained using single-pixel events, while resolutions obtained with polarization-correlated Compton events range from 3.6$\pm$0.3 mm to 4.9$\pm$0.3 mm, depending on data selection criteria. We also found that the polarization-correlated Compton events exhibit up to 20\% higher average signal to random background ratio compared to the single-pixel events. We also present the first imaging of the NEMA NU-4 phantom filled with a $^{68}$Ga solution of 378 MBq initial activity, successfully combining polarization-correlated events with conventional single-pixel event selection. Based on the extracted spatial resolution, signal-to-background, signal-to-noise, contrast, and contrast-to-noise ratio, we estimate that up to 10\% sensitivity increase may be attained by exploiting the polarization-correlated events, while preserving a high image quality.
\end{abstract}
\noindent\textbf{Keywords:} Positron Emission Tomography; quantum entanglement; photon polarization; NEMA NU-4 phantom; spatial resolution; image quality

\section*{Introduction}

Positron emission tomography (PET), an irreplaceable diagnostic tool for medical imaging, is based on the coincident detection of two gamma photons originating from positron annihilation in a patient's body. The gamma photons are typically detected by scintillation detectors capable of measuring photon energy, time of arrival, and hit position. PET has been continuously improved in terms of sensitivity and image quality by advancing both detector components and image reconstruction methods, leading to shorter acquisition times, using lower doses of radiopharmaceuticals, and setting the ground for imaging radiotracer kinetics\cite{Meikle2021, Schaart2021}.

In the past decade, the correlation of the polarizations of the gamma rays from positron annihilation has gained interest, initiated by simulation studies suggesting a possibility of enhancing PET \cite{Kuncic2011, Mcnamara2014, Toghyani2016, Watts2021}. 
This correlation is a consequence of quantum mechanical properties of the $\mathrm{e^+e^-}$ system. Conservation of angular momentum and parity following its decay, requires the polarizations of the annihilation quanta to be orthogonal\cite{dirac1928}. Moreover, it was postulated \cite{pryce, Snyder1948} and confirmed experimentally \cite{Wu1950, Langhoff1960, Kasday1975} that the annihilation quanta represent a quantum-entangled system\cite{Bohm1957}. This property, not previously utilized in PET, presents an additional opportunity independent of energy criteria, to discriminate true coincidences from random coincidence events, where no correlation is present.

The method to determine the polarization plane of a gamma photon is to let it undergo Compton scattering and measure its azimuthal scattering angle, which is related to the initial polarization plane. In the particular case of the back-to-back gamma rays from positron annihilation, where both of them undergo Compton scattering, the joint cross-section is given by \cite{pryce},  
\begin{equation}
\mathrm{ \frac{d^2\sigma}{d\Omega_1\!d\Omega_2}\!\!=\!\!\frac{r_0^4}{16}\left[F(\theta_1\!)F(\theta_2\!)-\!G(\theta_1\!)G(\theta_2\!)cos(2(\phi_1-\phi_2))\right] }
\label{DDCS}
\end{equation}
where $\mathrm{r_0}$ is the classical electron radius, $\mathrm{d\Omega_{1,2}}$ are the solid angles, $\mathrm{\theta_{1,2}}$ are the Compton scattering angles, and $\phi_1$ and $\phi_2$ are the azimuthal scattering angles of photon 1 and 2, respectively. The latter of the kinematic factors\cite{pryce} $\mathrm{F(\theta_{1,2})\!=\!\frac{2+(1-cos\theta_{1,2})^3}{(2-cos\theta_{1,2})^3}}$ and $\mathrm{G(\theta_{1,2})\!=\!\frac{sin^2\theta_{1,2}}{(2-cos\theta_{1,2})^2}}$ can be identified as the analyzing powers of polarimeters 1 and 2, respectively, and their product:
\begin{equation}
    \mu=\mathrm{G(\theta_{1})G(\theta_{2})}
\end{equation}
is labeled the modulation factor.

Since the measurement of polarization correlations of annihilation quanta is not currently utilized in clinical imaging with PET, the key question is how to implement it cost-effectively. Compton polarimeters need to determine the energy and position of the recoil electron and the scattered photon, which is typically achieved using two position-sensitive detector layers. However, using multiple detector layers on a large scale, as in the case of a clinical PET camera,  would considerably increase the complexity and cost of the apparatus. To avoid this, we implemented the concept of a single-layer detector in which a single segmented scintillator matrix is used as the gamma-ray polarimeter, acting as both the Compton scatterer and the absorber. The feasibility of the concept to measure polarization correlation of annihilation quanta was previously demonstrated under laboratory conditions\cite{Makek2020}, and further optimized\cite{Kozuljevic2021, Parashari2022_nima}. A similar crystal-based concept investigated the utilization of polarization correlations for PET imaging under laboratory conditions \cite{Kim2023}. Other approaches to include quantum-entangled Compton events in PET imaging are under investigation using plastic scintillators \cite{Moskal2024b}, and concepts based on Cadmium-Zinc-Telluride detectors \cite{Romanchek2024}.

This work reports the performance of a demonstrational setup, based on the single-layer concept for implementing polarization measurement in PET, and for the first time, presents the corresponding PET images of source distributions with clinically relevant activities, obtained by utilizing polarization-correlated Compton events. Further, we compare these to images obtained by conventional single-pixel event selection in terms of spatial resolution, signal-to-background, signal-to-noise, contrast, and contrast-to-noise, and estimate a sensitivity gain by exploiting the polarization-correlated inter-crystal scatter events. 

\section*{Methods}
\subsection*{Experimental setup}

\begin{figure}
    \centering
    \includegraphics[width=0.35\linewidth]{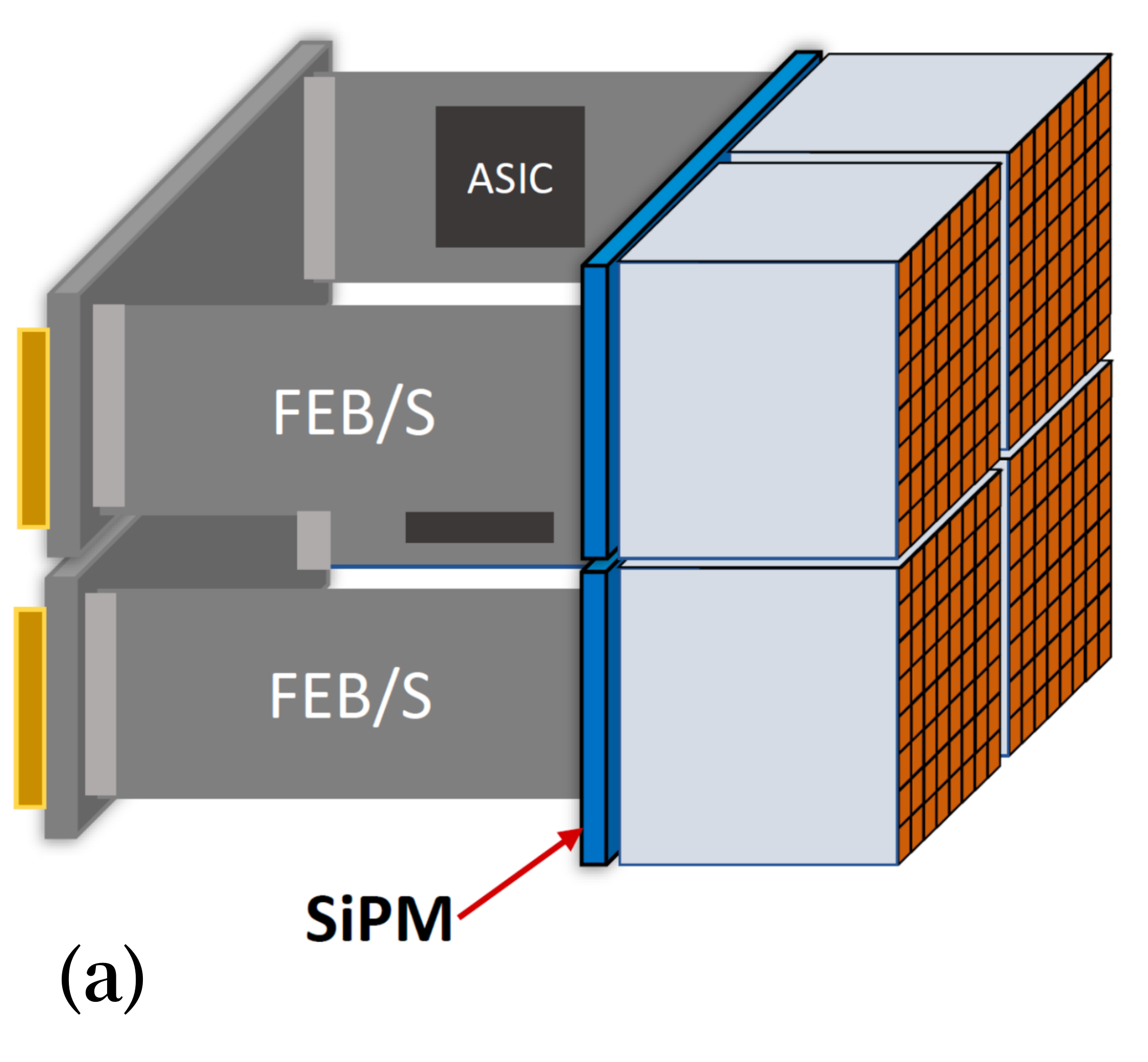} \hspace{5mm}
    \includegraphics[width=0.48\linewidth]{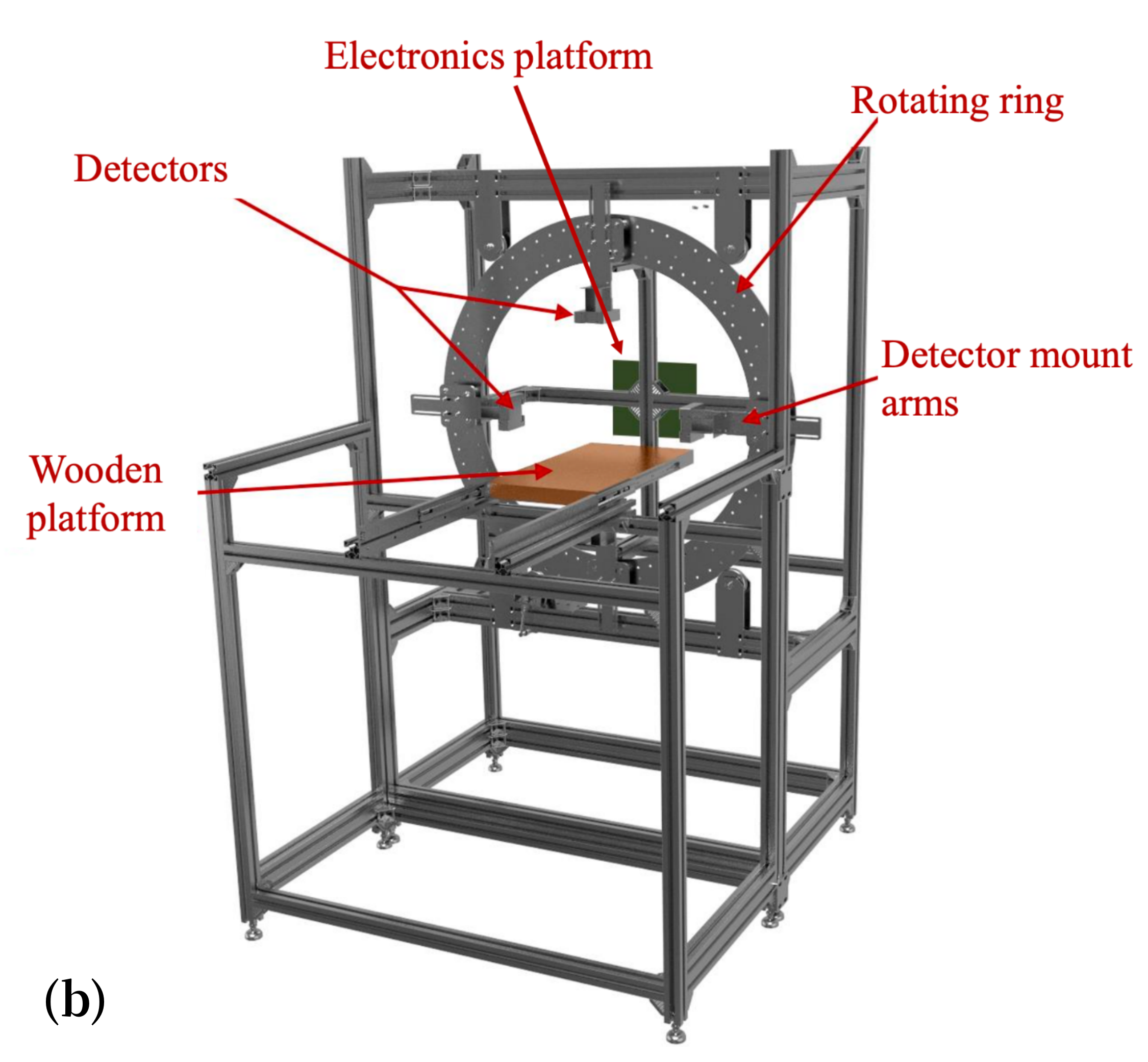} 
    \caption{Schematic illustration of (a) one detector module consisting of four $8\times 8$ crystal matrices, SiPMs, and front-end boards and (b) the experimental setup.} 
    \label{fig:module}
\end{figure}

The experimental setup comprised two detector modules based on segmented scintillators coupled to silicon photomultiplier (SiPM) arrays and front-end readout electronics (Figure \ref{fig:module} (a)). Each module had a dedicated cooling device that served a dual purpose of venting the ASICs in the front-end boards and of maintaining stability of the SiPMs' temperature. The modules were mounted on an aluminum ring supported by mechanical construction, with a sliding platform to position the investigated object in the field of view (Figure \ref{fig:module} (b)). Each module contained 256 GAGG:Ce scintillator crystals with 3.2 mm pitch, assembled from four $8\times 8$ detector matrices. The length of the crystals was 20 mm, and their cross sections were 3.0 mm x 3.0 mm in one and 2.9 mm x 2.9 mm in the other module. The $8\times 8$ matrices were read out by SiPM arrays (Hamamatsu Photonics, model S13361-3050AE-08), with one pixel of the detector matching one in the SiPM. All modules were tested in the laboratory using a point source as reported by Parashari et al.\cite{Parashari2022_nima}. The modules were placed on opposite sides of the ring so they could collect annihilation photons in coincidence. The mean energy resolution of the modules at 511 keV was found to be $(9.3\pm2.2)$\% and $(8.1\pm1.1)$\%, respectively \cite{Parashari2022_nima}. 

The mechanical support structure was engineered to provide a reliable foundation necessary for the imaging process, as well as to enable future upgrades of the system (Figure \ref{fig:module} (b)). It features the gantry, which rotates during the scan to emulate a full ring of detectors. The gantry is supported by four polytetrafluoroethylene (PTFE) rollers and an additional roller is connected to a stepper motor, ensuring precise rotation. The objects (sources) are placed on the platform, which can slide into the field of view. The detector arms mounted on the gantry allow fine-tuning of the module position, and they can slide to allow the diameter of the scanner from 400 to 700 mm. In the present work, the imaging was performed at two diameters, 620 mm and 430 mm. The first was chosen to assess the spatial resolution of the system at diameters similar to those of clinical scanners. The second, smaller diameter was selected to test the performance of the system at higher counting rates. 

\subsection*{Data acquisition and analysis}
\label{daq}
Data acquisition (DAQ) was performed by the TOFPET2 system (PETsys Electronics SA, Portugal)\cite{Francesco2016}. It was set to trigger on coincidence events with at least one pixel firing in each of the modules placed on the opposite sides of the ring. The DAQ process was automated in such a way that the acquisition at one position was followed by a precise rotation to the next position. For the 620 mm (430 mm) diameter, the data were acquired at 16 (12) different angular positions, each rotated by $11.25^\circ$ ($15.00^\circ$). 

From the full data set, we selected events where both gamma photons undergo Compton scattering within the respective module, by the following requirements: (a) exactly two pixels fire ($\mathrm{E_{px}}>120$ keV) in a module, (b) the energy of the 511 keV gamma is shared between these two pixels, requiring their sum to be within $\pm3\sigma$ of the 511 keV peak maximum, (c) energy depositions correspond to Compton scattering kinematics, as described by Parashari \textit{et al.} 2022\cite{Parashari2022_nima}. The energy threshold of 120 keV per pixel was selected to confidently reject the counts caused by the electronic noise. This threshold rejects the Compton scattering events with $\theta$<45$^\circ$, a region of very low sensitivity for polarization correlations, and hence is not the focus of this study. In events satisfying these criteria, it is not possible to distinguish with certainty which pixel corresponds to the recoil electron and which to the Compton-scattered photon absorption. According to GEANT4\cite{AGOSTINELLI2003250} simulations of the setup\cite{Kozuljevic2021}, on average, a slight preference ($\approx51\%$ of cases) is given to the forward scattering scenario. However, specific event topologies depending on energy depositions and inter-pixel distance favor selecting the pixel with higher energy deposition as the impact point. Therefore, the Compton scattering angles were reconstructed using a simulation-guided decision on the first fired pixel depending on event topology:

\begin{equation}
     \theta \!=\! \mathrm{acos\left(\frac{m_ec^2}{E_{px_1}\!\!+\!E_{px_2}} \!-\! \frac{m_ec^2}{E_{px_2}}\!+1\!\right)}; \;\;\phi \!=\! \mathrm{atan}\left(\frac{\Delta y}{\Delta x}\right)
     \label{eqn:angles}
\end{equation}
Further event selection was done based on event topology, i.e., the distance of the fired pixels in the respective module, $d_{1,2}=\sqrt{\Delta x^2+\Delta y^2}$, where (x,y) are coordinates of the pixel centers in the module reference frame. 

\begin{figure}
    \centering
    \includegraphics[width=0.8\linewidth]{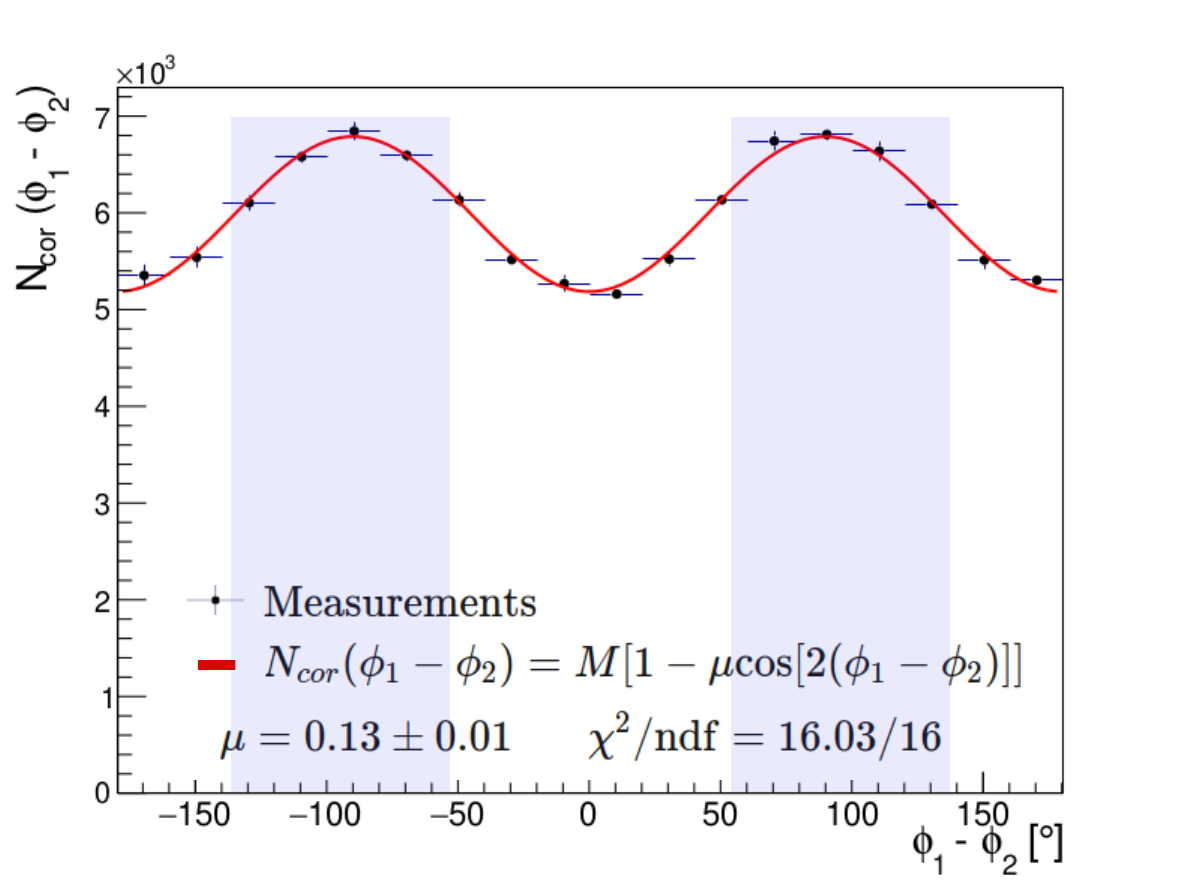}
    \caption{Acceptance-corrected azimuthal difference distribution, and the polarimetric sensitivity $\mu$ extracted by fitting eq. \ref{eqn:pw-fit}, for $60^\circ < \theta_{1,2} < 90^\circ$, and $3.1\; \mathrm{mm}<d_{1,2}<70 \; \mathrm{mm}$.}
    \label{fig:mu_plot}
\end{figure}

The detectors have non-uniform azimuthal acceptance ($\phi$), since the absorption probability of the scattered gamma photon depends on the distance traveled. Therefore, the $\phi$ angles covered by a combination of pixels with a large inter-pixel distance $d$, are less likely to be detected. To take this into account, the raw distribution $\mathrm{N (\phi_1 - \phi_2)}$ distribution, was acceptance-corrected according to:
\begin{equation}
    N_{cor}(\phi_1 - \phi_2)=\frac{N(\phi_1 - \phi_2)}{A_{mix}(\phi_1 - \phi_2)}
\end{equation}
where $A_{mix}(\phi_1 - \phi_2)$ is the $\phi_1-\phi_2$ distribution obtained by the event mixing technique\cite{Makek2020}. In this technique, the reconstructed angle $\phi_1$ of the gamma in one module is combined with the $\phi_2$ of the gamma in the opposite module, but from different events; thus they have no correlation, and the distribution is the result of detector pair acceptance. In this analysis, each single gamma event in one detector module was mixed with $10^2$ uncorrelated events in the other detector module to minimize the influence of the statistical error of $A_{mix}(\phi_1 - \phi_2)$ on $N_{cor}(\phi_1 - \phi_2)$.
This allowed us to extract the modulation factor, $\mu$, as a measure of the polarimetric sensitivity of the system, by fitting the $N_{cor}(\phi_1 - \phi_2)$ distribution with:
\begin{equation}
    N_{cor}(\phi_1 - \phi_2)=M [1 - \mu cos[2(\phi_1 - \phi_2)]] 
    \label{eqn:pw-fit}
\end{equation}
as demonstrated in Figure \ref{fig:mu_plot}. Finally, the event sample for image reconstruction was selected by requiring the Compton scattering angles $\theta_{min} < \theta_{1,2} < \theta_{max}$, and $|\phi_1-\phi_2|-90^\circ < \Delta\phi_{max}$.
The benchmark sample of single-pixel events (SPE) was selected by requiring a single hit between $435\;\mathrm{keV} < \mathrm{E_{px}} < 585\; \mathrm{keV}$ corresponding to each gamma photon. 

The TOFPET2 DAQ system recorded time stamps of individual hits; hence, the coincidence time was defined as the difference $\mathrm{t_1-t_2}$ between the time stamps of the hits in opposite modules. The true coincidence events (T) were selected within ±2.5 ns of the maximum, around zero. The random background sample (R) was selected by taking 2.5 ns windows on each side of the maximum, ranging from -12.5 to - 10 ns and 10 to 12.5 ns, respectively, as depicted in Figure \ref{fig:time_spectra_plots}. The random background sample (R) was selected by taking 2.5 ns windows on each side of the maximum. The signal-to-random background ratio was then defined as: $\mathrm{SBR=(T-R)/R}$. The coincidence time information was not used to deduce the Time-of-Flight, since this study focused on imaging objects in the center of the field of view.

\subsection*{Image reconstruction and assessment}
Image reconstruction was performed using the open-source software OMEGA\cite{omega} in MATLAB. Since the data were acquired with the detectors rotated around the source at a total of 16 (12) positions, for each position data were supplied to OMEGA as a separate module to emulate a full ring geometry with a total of 8192 crystals arranged in 16 trans-axial rings. The input data were prepared in event-by-event (list-mode) format containing the energy and the coordinates of the pixels fired under selected criteria. The lines of response (LOR) were formed using the coordinates of the centers of pixel faces, in which the first interaction was assumed to happen. The \textit{Ordered Subset Expectation Maximization} algorithm \cite{Hudson1994} was used for the image reconstruction. The number of iterations for each reconstruction was 10. The number of subsets was 8 for reconstruction of images obtained at 620 mm diameter, while for the images reconstructed from measurements at 420 mm was 6. The images do not have any additional reconstruction filters or enhancements applied. The field of view (FOV) was 128 mm $\times$ 128 mm $\times$ 51.2 mm with an image matrix of $96\times96\times31$ and $64\times64\times31$, for imaging of the rod sources and the NEMA-NU4 phantom, respectively.

To assess the image quality, in this work, we use several figures of merit:
\begin{itemize}

\item Signal to background, S/B = Intensity mean (signal) / Intensity mean (background),
\item Signal to noise, SNR = Intensity mean (signal) / Intensity standard deviation (signal), 
\item Contrast = [Intensity mean (hot) - Intensity mean (cold)] / Intensity mean (hot) $\times100$\%, 
\item Contrast-to-noise, CNR = [Intensity mean (hot) - Intensity mean (cold)] / Intensity standard deviation (hot).
\end{itemize}
where the statistical quantities run over all voxels in the selected region of interest (ROI). In what follows, we will compare the images reconstructed using polarization-correlated Compton events (PCE), with images based on events where each gamma hits only in a single pixel (SPE), as the benchmark. 

\begin{figure}
    \centering
    \includegraphics[width=0.7\linewidth]{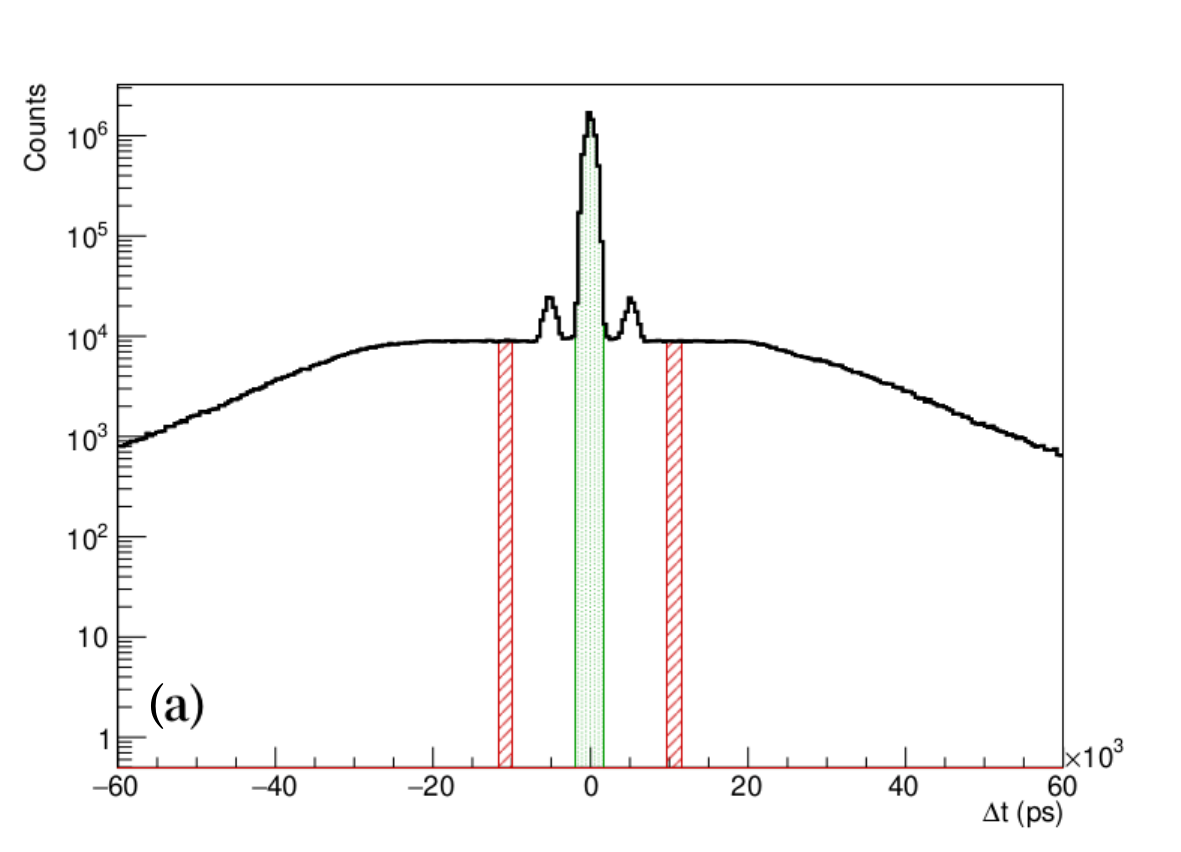}
    \includegraphics[width=0.7\linewidth]{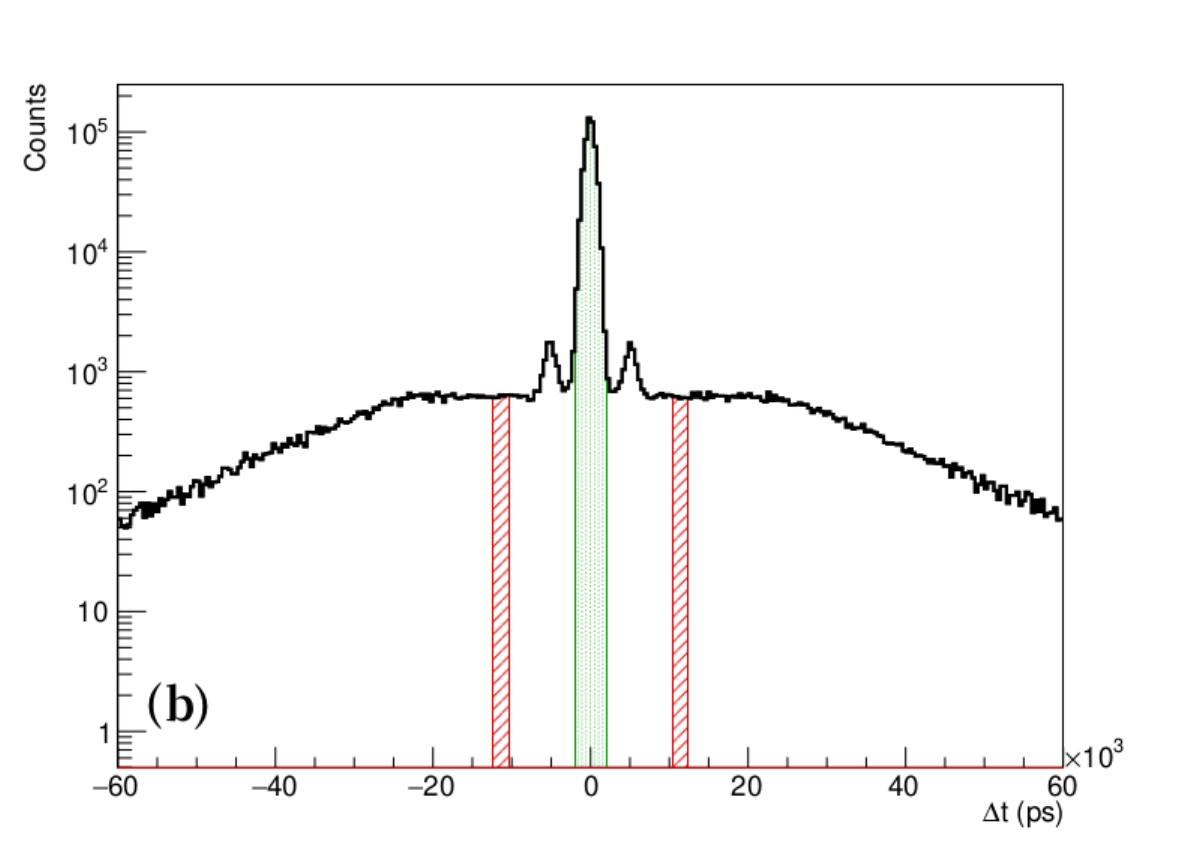}
    \caption{Coincidence time spectra for a) single-pixel events and for b) polarization-correlated Compton events. The signal window 5 ns wide is marked green, while the windows for random estimates 2.5 ns each are marked in red.}
    \label{fig:time_spectra_plots}
\end{figure}

Reconstructed images in the following sections were obtained after subtraction of random events by the delayed-window technique. The delayed window technique is a method of estimation and correction of random coincidences in PET scans \cite{knoll, brasse2005}. The technique relies on measurements of coincidences outside the coincidence window, as shown in Figure \ref{fig:time_spectra_plots}. The subtraction of the random coincidences was built in the reconstruction process in OMEGA software.

\section*{Results}
\label{Results}
Measurements with extended sources used for quality control of clinical or preclinical PET scanners were performed to test the system response at clinically relevant count rates comparable with patient imaging. The measurements were performed at the University Hospital Centre Zagreb.

\subsection*{Imaging with rod sources}
The first set of measurements involved two rod sources ($\mathrm{^{68}Ge,\; t_{1/2}\!=\!270\;days}$, active diameter 1.6 mm, overall diameter 3.2 mm, 45.5 MBq, each). The sources were positioned in the central field of view, parallel to the scanner axis, 2 cm apart, inside an epoxy-based (3D-printed) phantom. The acquisition time was 3 hours per position, with 16 (12) positions covering the full circle at 620 mm (430 mm) diameter.

\begin{figure}
    \centering
    \includegraphics[width=0.48\linewidth]{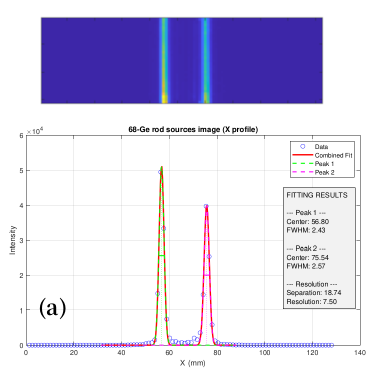}
    \includegraphics[width=0.48\linewidth]{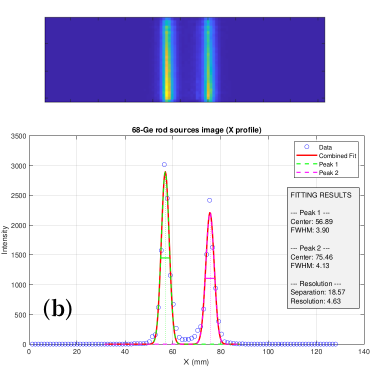}
    \caption{Reconstructed axial image of two $^{68}$Ge rod sources (31 slices), and the profile for central slices (8-23) using a) single-pixel events, and b) polarization-correlated Compton events selected by $45^\circ < \theta_{1,2} < 90^\circ$, $|\phi_1-\phi_2|-90^\circ < 45^\circ$, and $3.1\; \mathrm{mm}<d_{1,2}<10 \; \mathrm{mm}$ (6.5 mm), with ring diameter 430 mm.}
    \label{fig:line_sources_430}
\end{figure}

\begin{figure}
    \centering
    \includegraphics[width=0.48\linewidth]{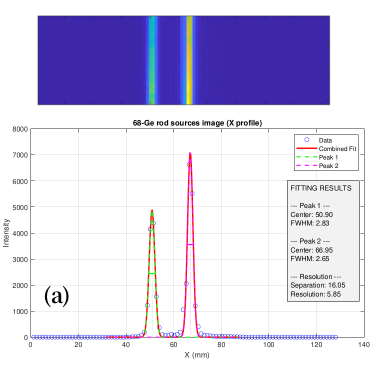}
    \includegraphics[width=0.48\linewidth]{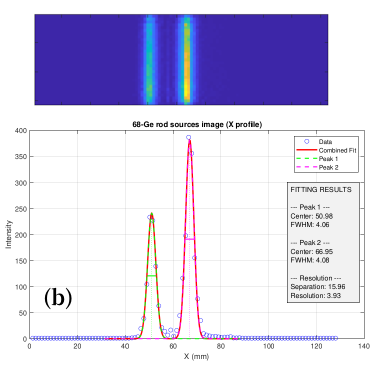}
    \caption{Reconstructed axial image of two $^{68}$Ge rod sources (31 slices), and the profile for central slices (8-23) using a) single-pixel events, and b)  polarization-correlated Compton events selected by $45^\circ < \theta_{1,2} < 90^\circ$, $|\phi_1-\phi_2|-90^\circ < 45^\circ$, and $3.1\; \mathrm{mm}<d_{1,2}<10 \; \mathrm{mm}$ (6.5 mm), with ring diameter 620 mm.}
    \label{fig:line_sources_620}
\end{figure}

For imaging at 430 mm ring diameter, the axial image of the rod sources reconstructed with SPE-only and PCE-only are shown in Figure \ref{fig:line_sources_430} (a) and (b), respectively. The spatial resolution is the average of the full width at half maximum of the Gaussian fits to the peaks. The obtained spatial resolutions, extracted from the profile are $2.5\pm 0.1$ mm, and $4.0\pm 0.1$ mm for the SPE and PCE, respectively. For imaging at 620 mm ring diameter, the images reconstructed with the same conditions are shown in Figure \ref{fig:line_sources_620}. The obtained spatial resolution, extracted from the profile are $2.7\pm 0.1$ mm, and $4.1\pm 0.1$ mm for SPE and PCE, respectively. The slight decrease in resolution notable in the PCE image with 620 mm ring diameter is expected due to the more pronounced non-collinearity effect at larger diameters.

The poorer spatial resolution obtained with PCE, is assumed to stem from the uncertainty in the assignment of the first fired pixel in Compton events\cite{Kozuljevic2024}. To further investigate it, we reconstructed images of the same data set, using different selection criteria, described in Table \ref{tab:cuts}. The table also lists the resulting spatial resolution, and the number of additional lines of response (LOR) by PCE as a fraction of SPE (LOR gain). 

\begin{figure}
    \centering
    \includegraphics[width=0.7\linewidth]{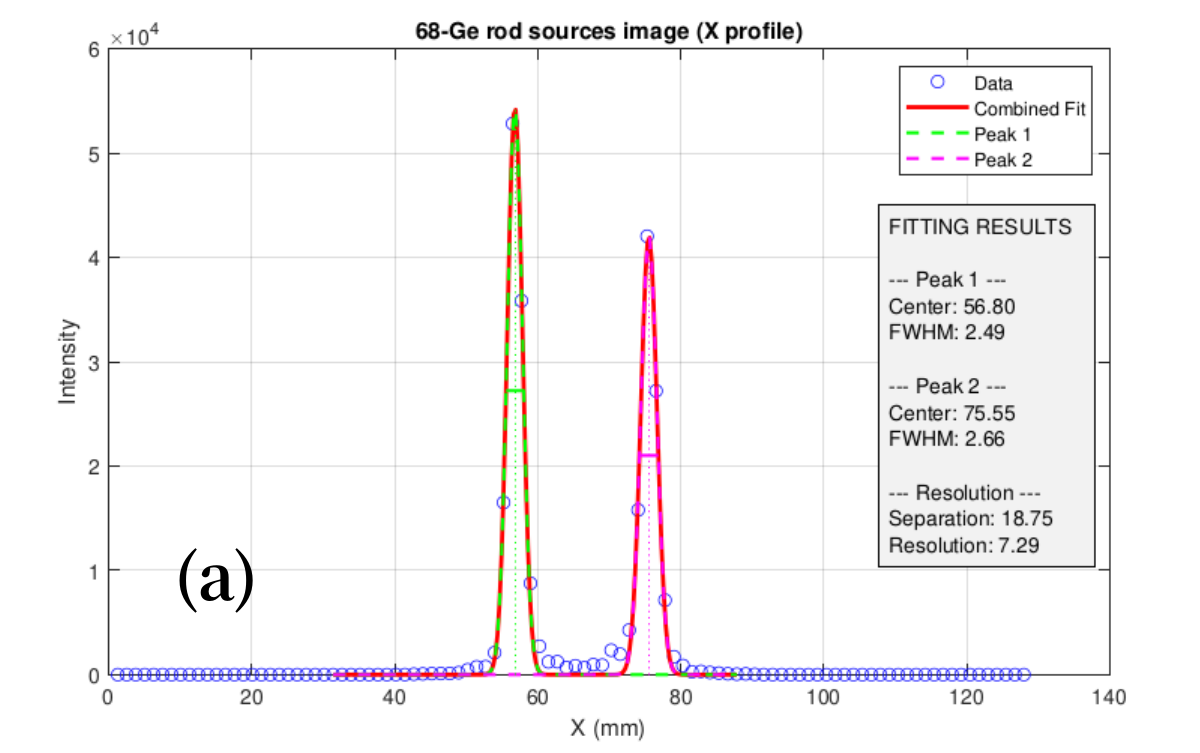}
    \includegraphics[width=0.7\linewidth]{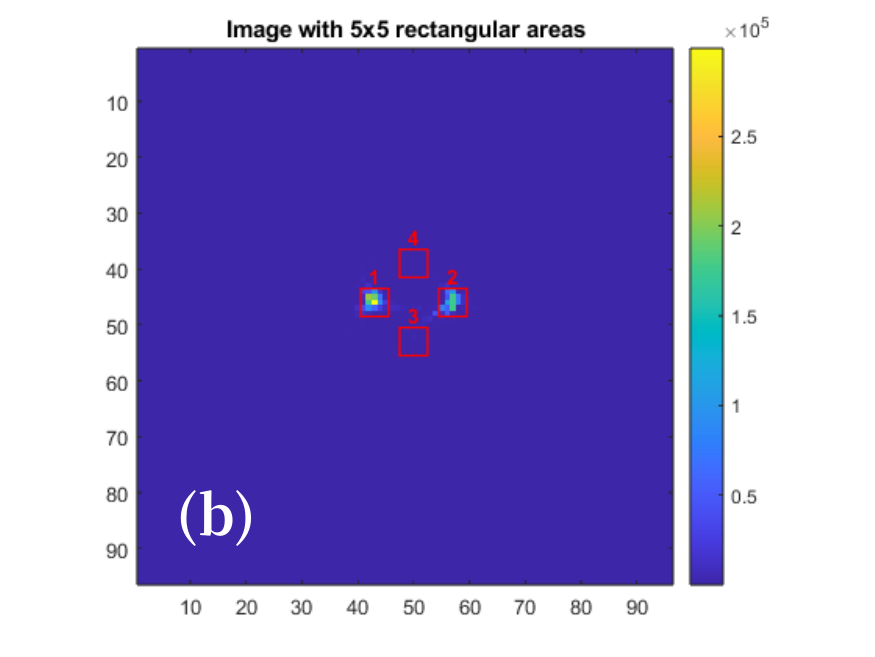}
    \caption{Reconstructed image of two $^{68}$Ge rod sources with merged SPE and PCE, with $45^\circ < \theta_{1,2} < 90^\circ$, $|\phi_1-\phi_2|-90^\circ < 45^\circ$, and 	$3.1\; \mathrm{mm}<d_{1,2}<10 \; \mathrm{mm}$ (6.5 mm): a) axial profile (slices 8-23), b) coronal view, with marked regions of interest.}
    \label{fig:merged_roi}
\end{figure}

\begin{figure}
    \centering
    \includegraphics[width=0.7\linewidth]{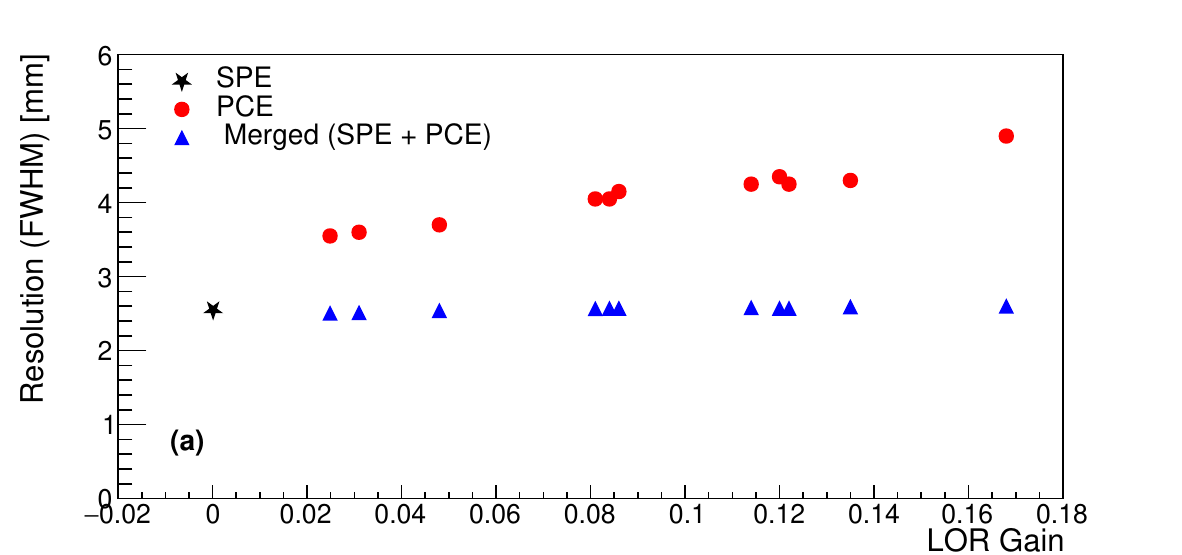}
    \includegraphics[width=0.7\linewidth]{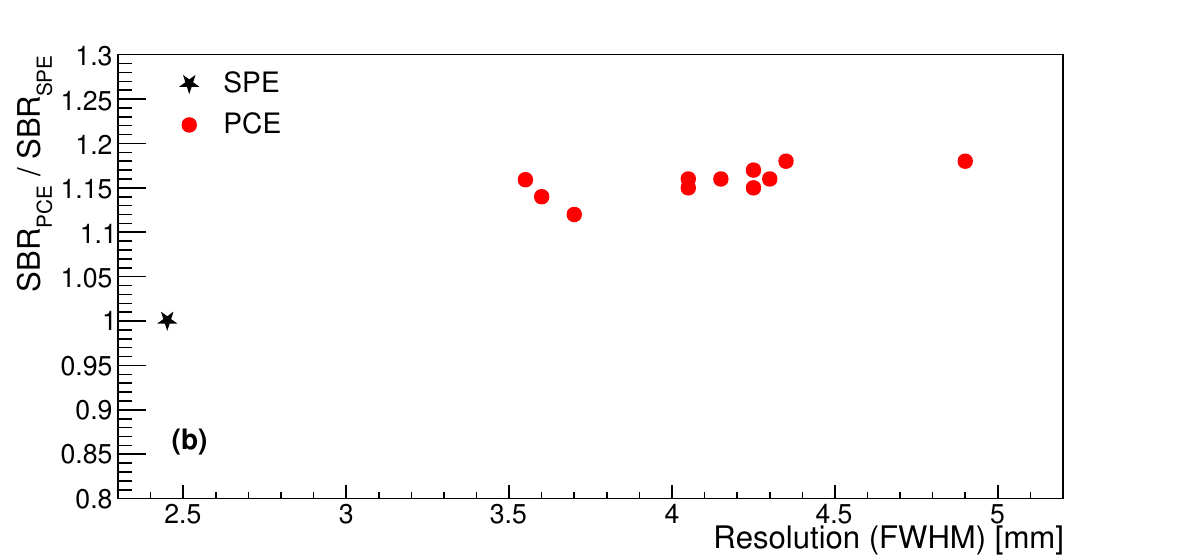}
    \caption{Spatial resolution of the SPE-image, PCE-only image and merged image vs. LOR gain (a). The uncertainties of determined resolutions in (a) are of the order of $\pm$ 0.1 mm. SBR of PCE over SBR of SPE vs. spatial resolution of the PCE-only image (corresponding uncertainties <5\%) (b). The data are from $\mathrm{^{68}Ge}$ rod source images for a ring diameter of 430 mm. }  
    \label{fig:spatial_vs_lor_430}
\end{figure}

Finally, we reconstructed image by merging SPE and PCE in list-mode (Figure \ref{fig:merged_roi} (a)), for different selection criteria of the latter. The spatial resolutions of these images are listed in the left-most column of the Table \ref{tab:cuts}, and the results demonstrate that they only marginally differ from the SPE image resolution, also shown in Figure \ref{fig:spatial_vs_lor_430} (a). In Figure \ref{fig:spatial_vs_lor_430} (b), we show the ratio of the SBR for PCE and for SPE extracted from the coincidence time plot (Figure \ref{fig:time_spectra_plots}, showing that selected PCE always contain a lower random fraction than the single-pixel events. 

\begin{figure}
    \centering
    \includegraphics[width=0.7\linewidth]{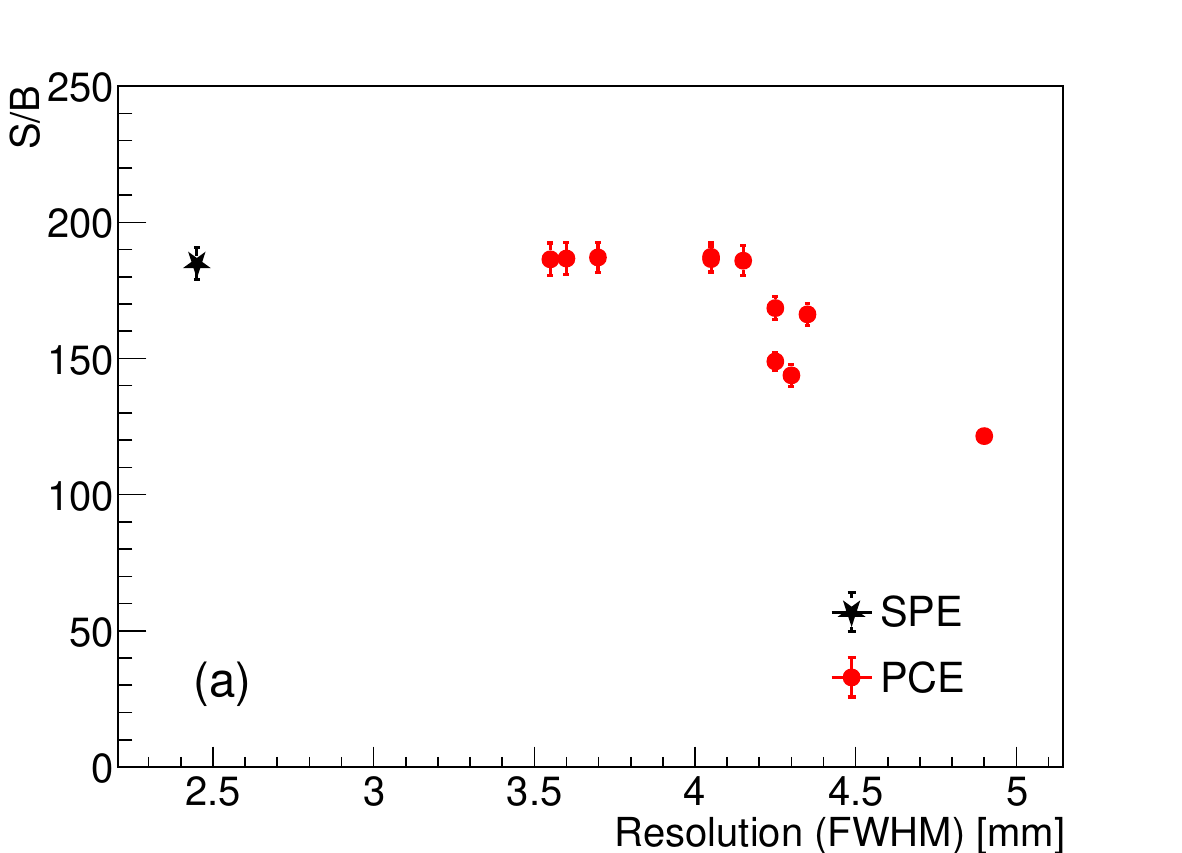}
    \includegraphics[width=0.7\linewidth]{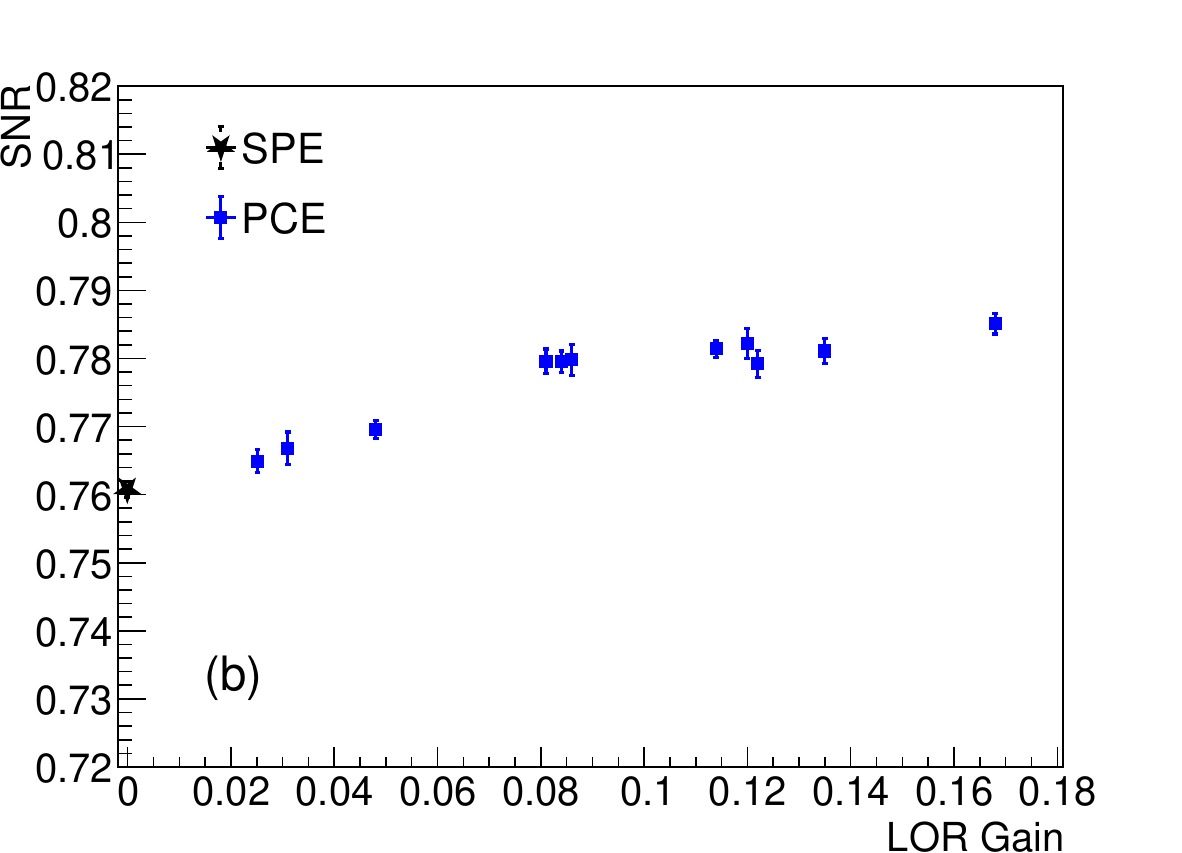}
    \caption{S/B of the merged SPE+PCE image vs spatial resolution of the PCE-only image (a). SNR of the merged SPE+PCE image vs LOR gain (b). The SPE points are marked for reference. The data are from $\mathrm{^{68}Ge}$ rod sources image for ring diameter 430 mm.}
    \label{fig:SB_SNR}
\end{figure}

The regions of interest for S/B and SNR evaluation are displayed in Figure \ref{fig:merged_roi} (b), where ROIs 1 and 2 correspond to signal, and ROIs 3 and 4 to background. These values are further evaluated as a function of LOR gain, and of the spatial resolution of PCE-only image in Figure \ref{fig:SB_SNR}, for different PCE selection criteria. To evaluate these figures of merit with a high statistical significance, we divide the full data set (list-mode) into 10 independent subsets, reconstruct the image for and extract the corresponding figure of merit. We finally report (Figures \ref{fig:SB_SNR} and \ref{fig:RC_vs_spatial_430}) the mean and the standard deviation across these 10 subsets for each figure of merit.

\begin{table}
\centering
\begin{threeparttable}
\caption{Various event selection criteria and corresponding SBR ratio in PCE over SPE, PCE LOR gain, spatial resolution of PCE-only image, and merged spatial resolution of merged SPE+PCE image. The uncertainties of determined resolutions are of the order of $\pm$ 0.1 mm, while for the calculated SBR of PCE over SBR of SPE the uncertainties are <5\%. The data are from $\mathrm{^{68}Ge}$ rod sources for ring diameter 430 mm.}
\label{tab:pet_angular}
\begin{tabular}{ccccccc}
\hline\hline
$\theta_{1,2}$ range & $|\phi_1-\phi_2|$ & $\mathrm{d_{1,2}}$ range (mm)  & SBR      &  LOR gain & Res (mm) & Res (mm) \\
    &     &      & (PCE/SPE)&       & (PCE)     &  (SPE+PCE) \\ \hline\hline
$60^\circ - 90^\circ$ & $90^\circ \pm 20^\circ$ & $3.1  \! < \! d \! < \! 6.5$ & 1.17 & 0.025 & 3.55 & 2.52 \\

$60^\circ - 90^\circ$ & $90^\circ \pm 45^\circ$ & $3.1  \! < \! d \! < \! 6.5$ & 1.14 & 0.031 & 3.60 & 2.52 \\

\hline 
\multirow{9}{4em}{$45^\circ - 90^\circ$} & \multirow{9}{4em}{$90^\circ \pm 45^\circ$} & $3.1  \! < \! d \! < \! 6.5$ & 1.12 & 0.048 & 3.70 & 2.55 \\
 &  & $3.1  \! < \! d \! < \! 10 (4.6)\tnote{*}$ & 1.15 & 0.081 & 4.05 & 2.57 \\
 &  & $3.1  \! < \! d \! < \! 10 (6.5)\tnote{*}$ & 1.16 & 0.084 & 4.05 & 2.58 \\
 &  & $3.1  \! < \! d \! < \! 10 (6.5)\tnote{*}$ & 1.16 & 0.086 & 4.15 & 2.58 \\
 &  & $3.1  \! < \! d \! < \! 15 (6.5)\tnote{*}$ & 1.17 & 0.114 & 4.25 & 2.59 \\
 &  & $3.1  \! < \! d \! < \! 15 (6.5)\tnote{*}$ & 1.18 & 0.120 & 4.35 & 2.58 \\
 &  & $3.1  \! < \! d \! < \! 70 (4.6)\tnote{*}$ & 1.15 & 0.122 & 4.25 & 2.58 \\
 &  & $3.1  \! < \! d \! < \! 70 (6.5)\tnote{*}$ & 1.16 & 0.135 & 4.30 & 2.60 \\
 &  & $3.1  \! < \! d \! < \! 70$       & 1.18 & 0.168 & 4.90 & 2.61 \\ \hline 
 Single-pixel & events & -- &  -- & -- & 2.5 \\ 
\hline\hline
\end{tabular}
\begin{tablenotes}
\item[*] In parenthesis is the maximum $\mathrm{d}$ required for at least one of the gamma photons.The value in parenthesis is the upper limit on inter-pixel distance d required for at least one of the gamma photons. LOR-to-pixel assignment depending on event topology is part of a separate study and will be published elsewhere.
\end{tablenotes}
\label{tab:cuts}
\end{threeparttable}
\end{table}

\subsection*{Imaging with NEMA-NU4 phantom}

\begin{figure}
    \centering
    \includegraphics[width=0.7\linewidth]{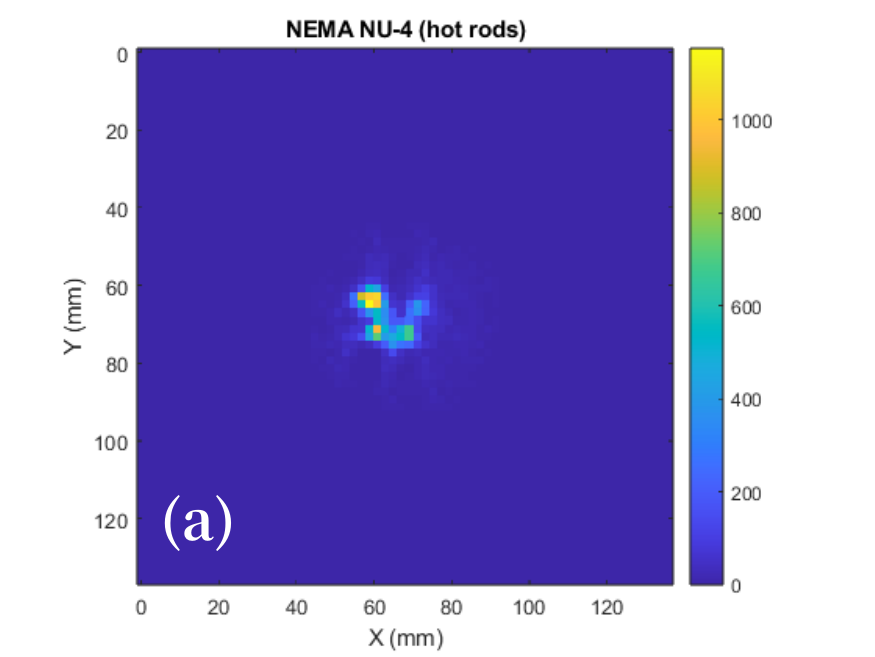}
    \includegraphics[width=0.7\linewidth]{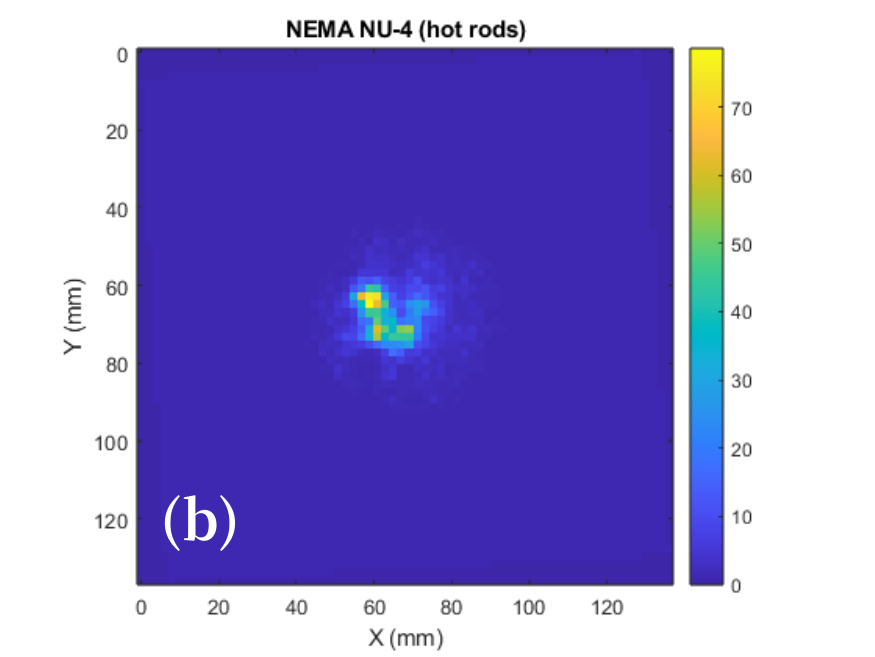}
    \includegraphics[width=0.7\linewidth]{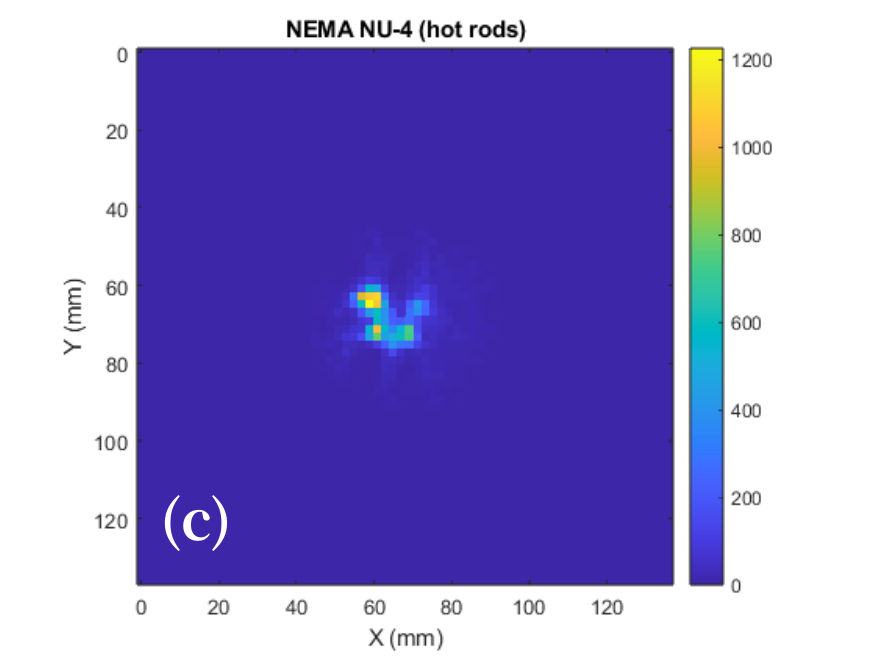}
    \caption{Reconstructed image of the NEMA-NU4 phantom (slices 3-5) with (a) SPE, (b) PCE, and (c) merged SPE+PCE. }
    \label{fig:NEMA_hot}
\end{figure}

\begin{figure}
    \centering
    \includegraphics[width=0.7\linewidth]{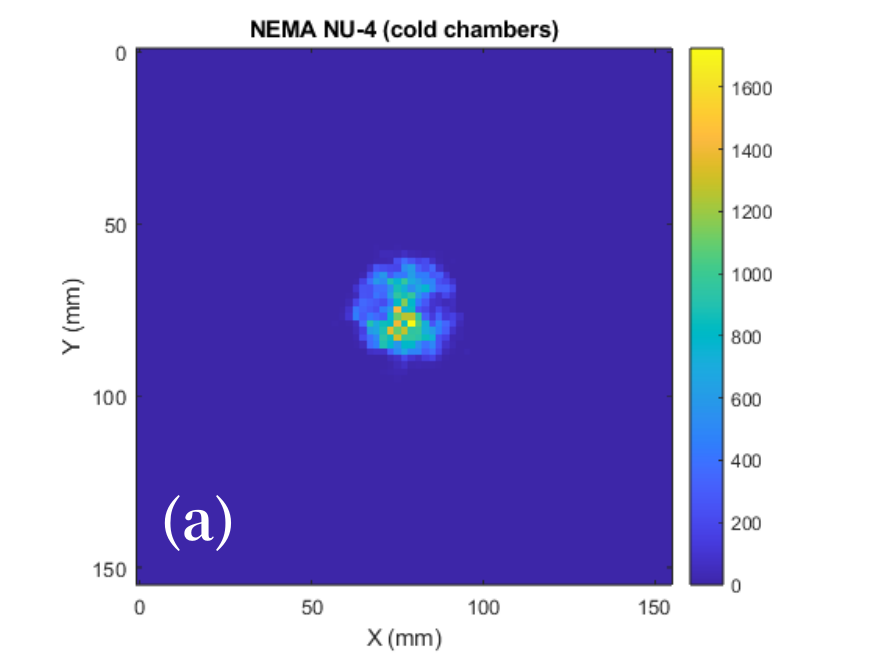}
    \includegraphics[width=0.7\linewidth]{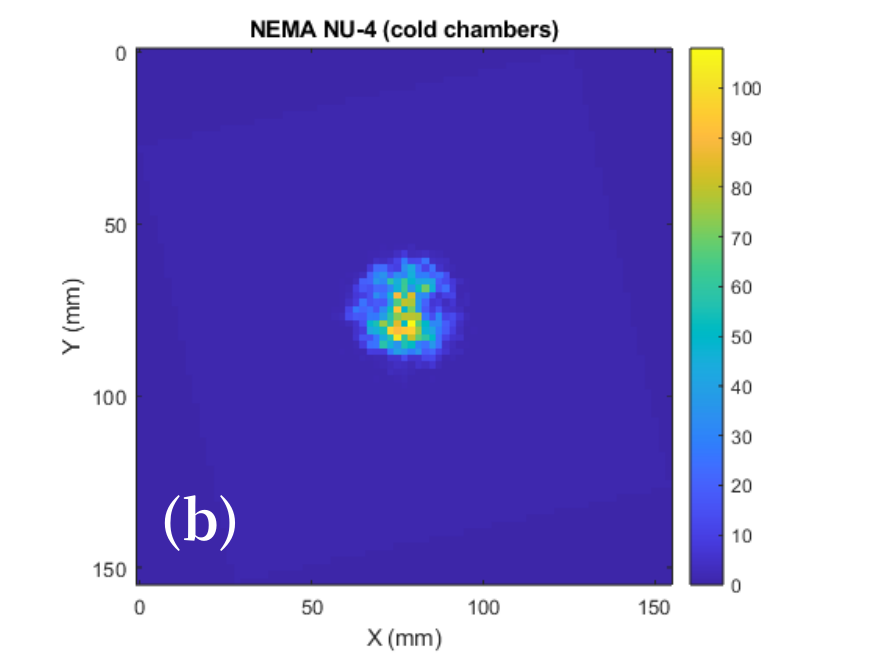}
    \includegraphics[width=0.7\linewidth]{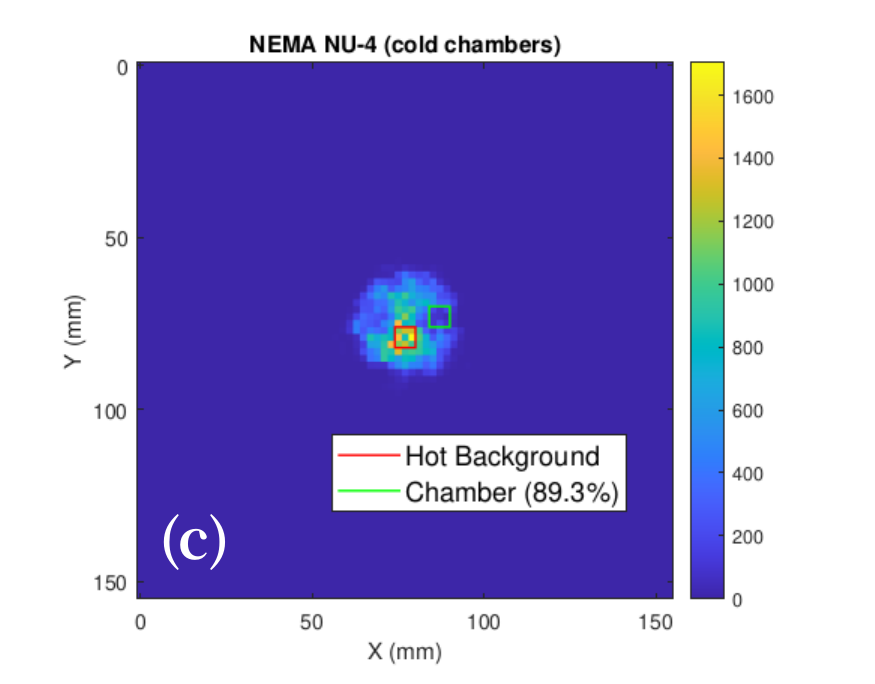}
    \caption{Reconstructed image of the NEMA-NU4 phantom (slices 20-26) with (a) SPE, (b) PCE, and (c) merged SPE+PCE. Regions of interest for contrast and CNR calculation are marked. The cold region is selected to correspond to the cavity filled with non-radioactive water, while the hot region corresponds to the chamber filled with radioactive solution surrounding the cold cavity.}
    \label{fig:NEMA_cold}
\end{figure}

\begin{figure}
    \centering
    \includegraphics[width=0.7\linewidth]{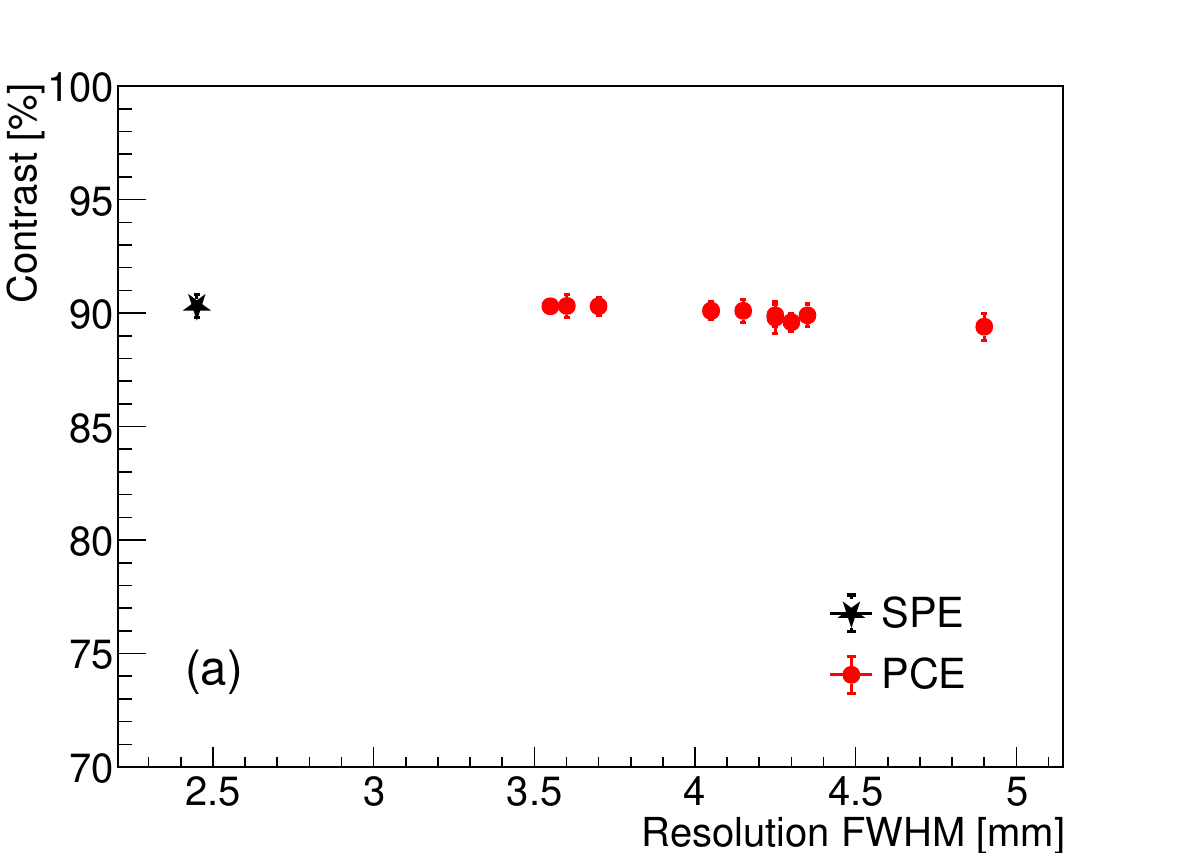} 
    \includegraphics[width=0.7\linewidth]{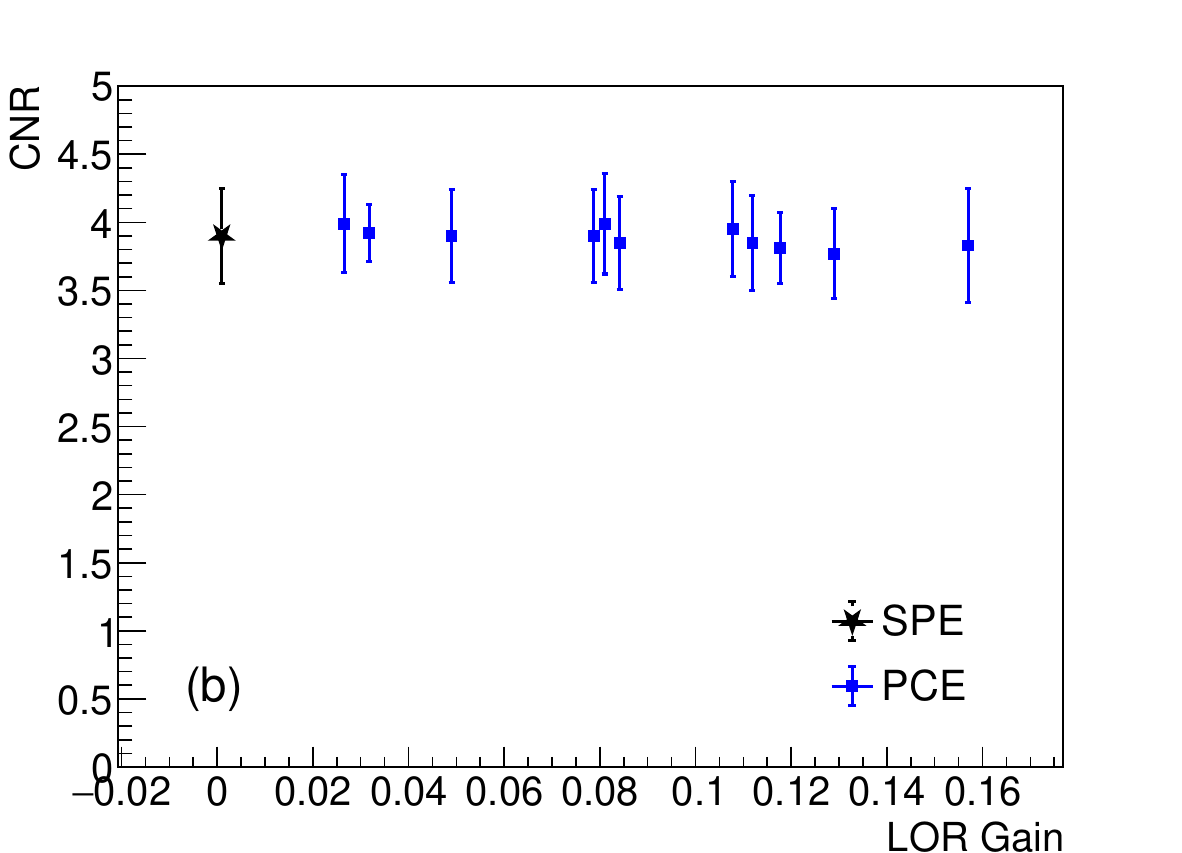} 
    \caption{Contrast in the merged SPE+PCE image for varying PCE-only spatial resolution (a). CNR in the merged SPE+PCE image vs. LOR gain (b). Data are for image of the NEMA-NU4 phantom of 378 MBq initial activity for ring diameter 430 mm.}
    \label{fig:RC_vs_spatial_430}
\end{figure}

In the second set of measurements, a NEMA-NU4 phantom filled with $^{68}$Ga solution of 378 MBq initial activity was imaged. The NEMA-NU4  phantom’s radioactive chamber consists of two parts: one where two cavities filled with non-radioactive water and air are submerged, and the other where the solution fills five rods (diameters of 1, 2, 3, 4 and 5 mm) radially aligned with the phantom length axis. The phantom was filled with solution of 68Ga eluate at the Radiochemistry Laboratory of the University Hospital Centre Zagreb, with immediate activity of 427 MBq. The measurement was started approximately 10 minutes after filling the phantom; therefore, the estimated initial activity of the source was 378$\pm$5 MBq. The measurements reported here are for the ring diameter of 430 mm, where the gantry was rotated to 12 adjacent positions to emulate the full ring. The acquisition time was adjusted to collect approximately the same number of events at each position, with the total scan duration of approximately 1h. Using the same criteria to select single-pixel events and polarization-correlated events as in the imaging of the rod sources described previously, we reconstructed images of the phantom. Figure \ref{fig:NEMA_hot} shows the part of the phantom with hot rods in the cold background, reconstructed with (a) SPE, (b) PCE, and (c) merged SPE+PCE, respectively. Figure \ref{fig:NEMA_cold} shows the part of the phantom with cold cylinders inside the hot background, reconstructed with (a) SPE, (b) PCE, and (c) merged SPE+PCE, respectively. In addition, we use the latter image to compare contrast and contrast-to-noise ratio for different event selection criteria. The contrast is shown vs. spatial resolution of PCE-only image in Figure \ref{fig:RC_vs_spatial_430} (a), where a notable correlation is visible. The CNR is shown as a function of LOR gain in Figure \ref{fig:RC_vs_spatial_430} (b).

\section*{Discussion}
PET imaging with polarization-correlated events was investigated in detail for the first time at clinically relevant activities. The PCE were used along with "standard" SPE in the list-mode to reconstruct the source image. These inter-crystal scatter events, which are mostly rejected by conventional PET scanners offer an additional resource, and the aim of the study was to assess the benefits of using them to improve image quality and/or to increase the scanner sensitivity. The detector technology used to perform this assessment, the single-layer scintillator matrix, is fully compatible with state-of-the-art clinical systems, underlying the relevance of the results for possible clinical translation.

One of the key results obtained, is the dependence of the spatial resolution of the PCE-generated image, on event selection criteria (Table \ref{tab:cuts}). This dependence is dominantly driven by the selected range of inter-pixel scatter distance ($\mathrm{d_{1,2}}$). The main reason for that is the uncertainty in the Compton interaction order, which results in uncertainty of assigning LOR to correct pixel position. Consequently, when the $\mathrm{d_{1,2}}$ range is smaller, this uncertainty is also lower, and the resulting resolution is better. It is worth noting that the spatial resolution of $2.5\pm 0.1$ mm (central FoV) obtained with SPE is compatible to state-of-art scanners, but also that the resolutions obtained with PCE, which are $<5$ mm (FWHM) are present in many installed clinical scanners\cite{Kolthammer2014, Rausch2015, Pan2019}. 

Another important aspect of the obtained results is that the PCE can add, depending on the selection criteria, up to 17\% of events for image reconstruction, of which up to 8\% events while preserving the highest image quality. It must be noted that the measurement was obtained with GAGG crystals with an $\mathrm{Z_{eff}}=54.5$, lower than LYSO ($\mathrm{Z_{eff}}=65$), so the corresponding fraction of inter-crystal Compton scatter is somewhat higher. A benefit of adding the polarization-correlated events is a lower contamination of the sample with randoms than the single-pixel events, as demonstrated in Figure \ref{fig:spatial_vs_lor_430} (b). 

Finally, the image quality parameters were assessed with sources of clinically relevant activities, comparing the image obtained with combined sample of SPE and PCE, with the SPE-only, as a benchmark. The results show that S/B ratio is maintained when using additional events, provided their spatial resolution is below 4 mm. At the same time SNR improves with additional PCEs (Figure \ref{fig:SB_SNR}). The contrast and CNR extracted from the images of the NEMA NU4 phantom do not show a pronounced dependence when adding polarization-correlated events, remaining within statistical uncertainties for all evaluated event selection criteria (Figure \ref{fig:RC_vs_spatial_430}). Since stricter event selection criteria improve spatial resolution, but reduce the number of additionally gained events, a trade-off between image resolution and sensitivity gain is necessary. Based on the extracted image parameters, we estimate that a sensitivity increase up to 10\% may be reached by exploiting the polarization-correlated events, while preserving a high image quality. Importantly, the results show guidelines on how to apply PCE to enhance PET imaging, pointing out the need to improve the LOR assignment procedure. There are several approaches that are under investigation, e.g. improved LOR assignment based on machine learning algorithms\cite{Zatcepin2020, Enriquez2024}, or application of depth-of-interaction capable detectors\cite{Petersen2024, Terragni2025}.  

\section*{Conclusion}
We have successfully constructed and operated a novel detector setup to demonstrate and assess the feasibility of using quantum-entanglement of the annihilation quanta in PET imaging by exploiting their pronounced polarization correlations. We demonstrated that these events have a reduced random fraction compared to conventional event selection. First images of sources with up to 378 MBq activities were successfully reconstructed using solely polarization-correlated events with spatial resolution below 5 mm in the central field of view. Further, we performed the first systematic quality assessment of the images obtained by merging the polarization-correlated events sample with the benchmark single-pixel events. While the reduced spatial resolution of the inter-crystal Compton event images, currently poses a barrier to achieve the full benefit of the quantum-entanglement enhanced PET imaging, the results still suggest that an improvement in sensitivity up to 10\% may be reached, while keeping or even slightly improving the image quality.

\bibliographystyle{unsrt}
\bibliography{bibliography_list}

\section*{Acknowledgements}
We are grateful to Mr. Marijan Žuvić of the University Hospital Centre Zagreb for the preparation of the radioactive source. 

\section*{Funding declaration}
This work was supported by the “Research Cooperability” Program of the Croatian Science Foundation, funded by the European Union from the European Social Fund under the Operational Programme Efficient Human Resources 2014–2020, grant number PZS-2019-02-5829. This work was also supported by the Croatian Science Foundation under the project number IP-2022-10-3878. S.P. acknowledges the financial support from the Severo Ochoa project CEX2023-001292-S funded by MICIU/AEI.

\section*{Author contributions statement}
M.M. conceived the experiments, secured the funding, and contributed to the manuscript, A.M.K. performed measurements, simulation, analysis, image reconstruction and contributed to the manuscript, T.B. participated in data taking and analysis, D.G. participated in data taking, Z.K. proposed the concept and secured the funding, S.P. performed measurements, contributed to the analysis and image reconstruction, L.P. contributed to the detector system development. All authors reviewed the manuscript.

\section*{Data availability}
The datasets used and/or analysed during the current study are available from the corresponding author upon reasonable request.

\end{document}